\documentclass[prd,showpacs,showkeys,nofootinbib,twocolumn]{revtex4-1}

\usepackage{amsmath,amsfonts,amssymb}

\begin{document}

\title{Equations of motion in metric-affine gravity: A covariant unified framework}

\author{Dirk Puetzfeld}
\email{dirk.puetzfeld@zarm.uni-bremen.de}
\homepage{http://puetzfeld.org}
\affiliation{ZARM, University of Bremen, Am Fallturm, 28359 Bremen, Germany} 

\author{Yuri N. Obukhov}
\email{obukhov@ibrae.ac.ru}
\affiliation{Theoretical Physics Laboratory, Nuclear Safety Institute, 
Russian Academy of Sciences, B.Tulskaya 52, 115191 Moscow, Russia}

\date{ \today}

\begin{abstract}
We derive the equations of motion of extended deformable bodies in metric-affine gravity. The conservation laws which follow from the invariance of the action under the general coordinate transformations are used as a starting point for the discussion of the dynamics of extended deformable test bodies. By means of a covariant approach, based on Synge's world function, we obtain the master equation of motion for an arbitrary system of coupled conserved currents. This unified framework is then applied to metric-affine gravity. We confirm and extend earlier findings; in particular, we once again demonstrate that it is only possible to detect the post-Riemannian spacetime geometry by ordinary (non-microstructured) test bodies if gravity is nonminimally coupled to matter.
\end{abstract}



\maketitle

We dedicate this article to Friedrich W.\ Hehl -- a pioneer of metric-affine gravity -- on the occasion of his birthday.

\section{Introduction}\label{introduction_sec}

Metric-affine gravity \cite{Hehl:1995} is a natural extension of Einstein's general relativity theory. It is based on gauge-theoretic principles \cite{Blagojevic:2002,Hehl:2013}, and it takes into account microstructural properties of matter (spin, dilation current, proper hypercharge) as possible physical sources of the gravitational field, on an equal footing with macroscopic properties (energy and momentum) of matter. 

In this work we derive the equations of motion of extended deformable test bodies in metric-affine gravity. In this theory, matter is characterized by three fundamental Noether currents -- the canonical energy-momentum current, the canonical hypermomentum current, and the metrical energy-momentum current. These objects satisfy a set of conservation laws (or, more exactly, balance equations). Following Mathisson, Papapetrou, and Dixon \cite{Mathisson:1937,Papapetrou:1951:3,Dixon:1964,Dixon:1974,Dixon:1979,Dixon:2008}, the equations of motion of extended test bodies are derived from the conservation laws. Our derivation is based on a covariant multipolar test body method, which utilizes Synge's world function formalism \cite{Synge:1960,DeWitt:Brehme:1960}. 

In view of the multi-current characterization of matter in metric-affine gravity, we develop here a general approach which is applicable to an arbitrary set of conservation laws for any number of currents. The latter can include the gravitational, electromagnetic, and other physical currents if they are relevant to the model under consideration. The results presented here allow for the systematic study of test body motion in a very large class of gravitational theories (and not only gravitational), in particular they can also be applied to the case in which there is a general nonminimal coupling between gravity and matter. Models with nonminimal coupling have recently attracted a lot of attention in the literature \cite{Bertolami:etal:2007,Nojiri:2011}.  Their physical interpretation and impact are still a subject of discussion \cite{Straumann:2008,Harko:2014:1}.

Here we explicitly show how the new geometrical structures in metric-affine gravity couple to matter, which in turn may underlie the design of experimental tests of gravity beyond the Einsteinian (purely Riemannian) geometrical picture. Our current work, generalizes and unifies several previous works \cite{Stoeger:Yasskin:1979,Stoeger:Yasskin:1980,Puetzfeld:2007,Puetzfeld:Obukhov:2008:1,Puetzfeld:Obukhov:2008:2,Puetzfeld:Obukhov:2013:1,Hehl:Obukhov:Puetzfeld:2013,Puetzfeld:Obukhov:2013:3,Puetzfeld:Obukhov:2013:4,Roshan:2013,Puetzfeld:Obukhov:2014:1} on the equations of motion in gauge gravity theories.

The structure of the paper is as follows: In section \ref{MAG_sec} we briefly introduce the relevant geometrical notions and recall the dynamical structure of metric-affine gravity. Our discussion is different from \cite{Hehl:1995} in that we avoid the use of the anholonomic frame/coframe, and all considerations are based on the traditional (Einsteinian) holonomic coordinate tensor formalism. We pay special attention to the extension of metric-affine gravity to the case of nonminimal coupling of gravity and matter. In section \ref{master_sec} we develop a generalized framework for the analysis of the multi-current conservation laws, and derive general covariant {\sl master equations} of motion for test bodies characterized by an arbitrary set of Noether currents. On the basis of these general results, we then obtain in section \ref{eom_sec} the equations of motion of extended test bodies in metric-affine gravity. The infinite hierarchy of equations for multipole moments up to an arbitrary order is given, and we analyze the lowest orders of approximation in some more detail. In particular we derive the equations of motion of a pole-dipole test body, as well as monopolar particle in section \ref{special_cases_sec}, and compare those to previous results in the literature. Our final conclusions are drawn our in \ref{conclusion_sec}. A brief summary of our conventions and frequently used formulas can be found in the appendices \ref{conventions_app} and \ref{expansion_app}. Appendix \ref{explicit_app} contains some supplementary material on the derivation of the general equations of motion.

Our notations and conventions are those of \cite{Hehl:1995}. In particular, the basic geometrical quantities such as the curvature, torsion, and nonmetricity are defined as in \cite{Hehl:1995}, and we use the Latin alphabet to label the spacetime coordinate indices. Furthermore, the metric has the signature $(+,-,-,-)$. It should be noted that our definition of the metrical energy-momentum tensor is different from the definition used in \cite{Bertolami:etal:2007,Nojiri:2011,Puetzfeld:Obukhov:2013:1}. 

\section{Metric-affine gravity}\label{MAG_sec}

The geometrical arena of metric-affine gravity is as follows. The physical spacetime is identified with a four-dimensional smooth manifold $L_4$, which is endowed with a metric $g_{ij}$, and a linear connection $\Gamma_{ki}{}^j$. These structures introduce the physically important notions of lengths, angles, and parallel transport on the spacetime. In general, the geometry of such a manifold is exhaustively characterized by three tensors: the curvature, the torsion and the nonmetricity. They are defined as follows
\begin{eqnarray}
R_{kli}{}^j &:=& \partial_k\Gamma_{li}{}^j - \partial_l\Gamma_{ki}{}^j + \Gamma_{kn}{}^j \Gamma_{li}{}^n - \Gamma_{ln}{}^j\Gamma_{ki}{}^n,\label{curv}\\
T_{kl}{}^i &:=& \Gamma_{kl}{}^i - \Gamma_{lk}{}^i,\label{tors}\\ \label{nonmet}
Q_{kij} &:=& -\,\nabla_kg_{ij} = - \partial_kg_{ij} + \Gamma_{ki}{}^lg_{lj} + \Gamma_{kj}{}^lg_{il}.
\end{eqnarray}
The Riemannian connection $\widehat{\Gamma}_{kj}{}^i$ is uniquely determined by the conditions of vanishing torsion and nonmetricity which yield explicitly 
\begin{equation}
\widehat{\Gamma}_{kj}{}^i = {\frac 12}g^{il}(\partial_jg_{kl} + \partial_kg_{lj} - \partial_lg_{kj}).\label{Chr}
\end{equation}
The deviation of the geometry from the Riemannian one is then conveniently described by the {\it distortion} tensor 
\begin{equation}
N_{kj}{}^i := \widehat{\Gamma}_{kj}{}^i - \Gamma_{kj}{}^i.\label{dist}
\end{equation}
The system (\ref{tors}) and (\ref{nonmet}) allows us to find the distortion tensor in terms of the torsion and nonmetricity. Explicitly,
\begin{eqnarray}
N_{kj}{}^i &=& -\,{\frac 12}(T_{kj}{}^i + T^i{}_{kj} + T^i{}_{jk})\nonumber\\
&& +\,{\frac 12}(Q^i{}_{kj} - Q_{kj}{}^i - Q_{jk}{}^i).\label{NTQ}
\end{eqnarray}
Conversely, one can use this to express the torsion and nonmetricity tensors in terms of the distortion,
\begin{eqnarray}
T_{kj}{}^i &=& -\,2N_{[kj]}{}^i,\label{TN}\\
Q_{kij} &=& -\,2N_{k(ij)}.\label{QN}
\end{eqnarray}
Substituting (\ref{dist}) into (\ref{curv}), we find the relation between the non-Riemannian and the Riemannian curvature tensors
\begin{equation}
R_{adc}{}^b = \widehat{R}_{adc}{}^b - \widehat{\nabla}_aN_{dc}{}^b + \widehat{\nabla}_dN_{ac}{}^b + N_{an}{}^bN_{dc}{}^n - N_{dn}{}^bN_{ac}{}^n.\label{RRN}
\end{equation}
The hat over a symbol denotes the Riemannian objects (such as the curvature tensor) and the Riemannian operators (such as the covariant derivative) constructed from the Christoffel symbols (\ref{Chr}).

\subsection{Dynamics in metric-affine theory}\label{fieldeqs_sec}

The gravitational effects in the metric-affine theory are described by the set of fundamental variables: the independent metric $g_{ij}$ and connection $\Gamma_{kj}{}^i$. Accordingly, there are two sets of field equations.

Assuming standard minimal coupling, the total Lagrangian of interacting gravitational and matter fields reads
\begin{equation}\label{Ltot}
L = V(g_{ij}, R_{ijk}{}^l, N_{ki}{}^j) + L_{\rm mat}(g_{ij}, \psi^A, \nabla_i\psi^A).
\end{equation}
In general, the gravitational Lagrangian $V$ is constructed as a diffeomorphism invariant function of the curvature, torsion, and nonmetricity. However, in view of the relations (\ref{TN}) and (\ref{QN}), we can limit ourselves to Lagrangian functions that depend arbitrarily on the curvature and the distortion tensors. The matter Lagrangian depends on the matter field $\psi^A$ and its covariant derivative $\nabla_k\psi^A = \partial_k\psi^A -\Gamma_{ki}{}^j\,(\sigma^A{}_B)_j{}^i\,\psi^B$. Here $(\sigma^A{}_B)_j{}^i$ are the generators of general coordinate transformations. 

The field equations of metric-affine gravity can be written in several equivalent ways. The standard form is the set of the so-called ``first'' and ``second'' field equations (using the modified covariant derivative defined by ${\stackrel * \nabla}{}_i = \nabla_i + N_{ki}{}^k$):
\begin{eqnarray}
{\stackrel * \nabla}{}_nH^{in}{}_k + {\frac 12}T_{mn}{}^iH^{mn}{}_k - E_k{}^i &=& - \Sigma_k{}^i,\label{1st}\\
{\stackrel * \nabla}{}_lH^{kli}{}_j + {\frac 12}T_{mn}{}^kH^{mni}{}_j - E^{ki}{}_j &=& \Delta^i{}_j{}^k.\label{2nd}  
\end{eqnarray}
Here the generalized gravitational field momenta are introduced by
\begin{eqnarray}
H^{kli}{}_j &:=& -\,2{\frac {\partial V}{\partial R_{kli}{}^j}},\label{HH}\\
H^{ki}{}_j &:=& -\,{\frac {\partial V}{\partial T_{ki}{}^j}},\label{HT}\\
M^{kij} &:=& -\,{\frac {\partial V}{\partial Q_{kij}}},\label{HM}
\end{eqnarray}
and the gravitational hypermomentum density is
\begin{equation}\label{EN}
E^{ki}{}_j = - H^{ki}{}_j - M^{ki}{}_j = -\,{\frac {\partial V}{\partial N_{ki}{}^j}}.
\end{equation}
Furthermore, the generalized energy-momentum tensor of the gravitational field is
\begin{equation}
E_k{}^i = \delta_k^i V + {\frac 12}Q_{kln} M^{iln} + T_{kl}{}^n H^{il}{}_n + R_{kln}{}^m H^{iln}{}_m.\label{Eg}
\end{equation}
The sources of the gravitational field are the canonical energy-momentum tensor and the canonical hypermomentum of matter, respectively:
\begin{eqnarray}
\Sigma_k{}^i &:=& {\frac {\partial L_{\rm mat}}{\partial\nabla_i\psi^A}}\,\nabla_k\psi^A - \delta^i_kL_{\rm mat}.\label{canD}\\
\Delta^i{}_j{}^k &:=& {\frac {\partial L_{\rm mat}}{\partial \Gamma_{ki}{}^j}} = - {\frac {\partial L_{\rm mat}}{\partial\nabla_k\psi^A}} \,(\sigma^A{}_B)_j{}^i \psi^B.\label{tD}
\end{eqnarray}

It is straightforward to verify that instead of the first field equation (\ref{1st}), one can use the so-called zeroth field equation which reads
\begin{equation}
{\frac 2{\sqrt{-g}}}{\frac {\delta (\sqrt{-g}V)}{\delta g_{ij}}} = t^{ij}.\label{0th}
\end{equation}
On the right-hand side, the matter source is now represented by the metrical energy-momentum tensor which is defined by
\begin{equation}\label{tmet}
t_{ij} := {\frac 2{\sqrt{-g}}}{\frac {\partial (\sqrt{-g}L_{\rm mat})}{\partial g^{ij}}}.
\end{equation}
The system (\ref{1st}) and (\ref{2nd}) is completely equivalent to the system (\ref{0th}) and (\ref{2nd}), and it is a matter of convenience which one is solved. 

In order to give an explicit example of physical matter with microstructure, we recall the hyperfluid model \cite{Obukhov:1993}. This is a direct generalization of the general relativistic ideal fluid variational theory \cite{Taub:1954,Schutz:1970} and of the spinning fluid model of Weyssenhoff and Raabe \cite{Weyssenhoff:1947,Obukhov:1987}. Using the variational principle for the hyperfluid \cite{Obukhov:1993}, one derives the canonical energy-momentum and hypermomentum tensors:
\begin{eqnarray}
\Sigma_k{}^i &=& \,v^iP_k - p\left(\delta_k^i - v^iv_k\right),\label{hypS}\\
\Delta^n{}_m{}^i &=& \,v^iJ_m{}^n,\label{hypD}
\end{eqnarray}
where $v^i$ is the 4-velocity of the fluid and $p$ is the pressure. Fluid elements are characterized by their microstructural properties: the momentum density $P_k$ and the intrinsic hypermomentum density $J_m{}^n$. 

\subsection{Nonminimal coupling}\label{nonmin_sec}

Let us now consider an extension of the metric-affine theory by allowing the {\it nonminimal coupling} of matter and gravity via the modified Lagrangian
\begin{equation}\label{Lnon}
FL_{\rm mat}(g_{ij}, \psi^A, \nabla_i\psi^A).
\end{equation}
which replaces the second term in (\ref{Ltot}). The coupling function $F = F(g_{ij}, R_{ijk}{}^l, N_{ki}{}^j)$ can depend arbitrarily on its arguments. When $F = 1$, we recover the minimal coupling case. 

In the previous paper \cite{Obukhov:Puetzfeld:2014} we derived the conservation laws in such a generalized theory. They read as follows:
\begin{eqnarray}\label{cons1f}
\widehat{\nabla}_j\Delta^i{}_k{}^j &=& -\,U_{jm}{}^{ni}{}_k\Delta^m{}_n{}^j + \Sigma_k{}^i - t_k{}^i,\\
\widehat{\nabla}_j\Sigma_k{}^j &=& -\,V_{j}{}^n{}_k\Sigma_n{}^j - R_{kjm}{}^n\Delta^m{}_n{}^j - {\frac 12}Q_{kj}{}^nt_n{}^j \nonumber \\
&&- A_k\,L_{\rm mat}.\label{cons2f}
\end{eqnarray}
Here we denote $A_k := \widehat{\nabla}_k\log F$, and
\begin{eqnarray}
U_{jmn}{}^{ik} &=& A_j\delta_m^i\delta_n^k - N_{jm}{}^i\delta_n^k + N_{j}{}^k{}_n\delta_m^i,\label{U}\\
V_{jn}{}^k &=& A_j\delta_n^k + N^k{}_{jn}.\label{V}
\end{eqnarray}

\section{General multipolar framework}\label{master_sec}

In this section we derive ``master equations of motion'' for a general extended test body, which is characterized by a set of currents 
\begin{equation}
J^{Aj}.\label{JA}
\end{equation}
Normally, these are the so-called Noether currents that correspond to an invariance of the action under certain symmetry group. However, this is not necessary, and any set of currents is formally allowed. We call $J^{Aj}$ dynamical currents. The generalized index (capital Latin letters $A,B,\dots$) labels different components of the currents. 

As the starting point for derivation of the equations of motion for generalized multipole moments, we consider the following conservation law:
\begin{equation}
\widehat{\nabla}_jJ^{Aj} = -\,\Lambda_{jB}{}^A\,J^{Bj} - \Pi^A{}_{\dot{B}}\Xi^{\dot{B}}.\label{dJ}
\end{equation}
On the right-hand side, we introduce objects that can be called material currents 
\begin{equation}
\Xi^{\dot{A}}\label{Xi}
\end{equation}
to distinguish them from the dynamical currents $J^{Aj}$. The number of components of the dynamical and material currents is different; hence, we use a different index with a dot, $\dot{A}, \dot{B}, \dots$,  the range of which does not coincide with that of $A,B,\dots$. At this stage we do not specify the ranges of both types of indices, this will be done for the particular examples which we analyze later. As usual, Einstein's summation rule over repeated indices is assumed for the generalized indices as well as for coordinate indices. 

Both sets of currents $J^{Aj}$ and $\Xi^{\dot{A}}$ are constructed from the variables that describe the structure and the properties of matter inside the body. In contrast, the objects 
\begin{equation}
\Lambda_{jB}{}^A,\qquad \Pi^A{}_{\dot{B}},
\end{equation}
do not depend on the matter, but they are functions of the external classical fields which act on the body and thereby determine its motion. The list of such external fields includes the electromagnetic, gravitational, and scalar fields.

We will now derive the equations of motion of a test body by utilizing the covariant expansion method of Synge \cite{Synge:1960}. For this we need the following auxiliary formula for the absolute derivative of the integral of an arbitrary bitensor density $\widetilde{B}^{x_1 y_1}=\widetilde{B}^{x_1 y_1}(x,y)$ (the latter is a tensorial function of two spacetime points):
\begin{eqnarray}
{\frac{D}{ds}} \int\limits_{\Sigma(s)} \widetilde{B}^{x_1 y_1} d \Sigma_{x_1} &=& \int\limits_{\Sigma(s)} \widehat{\nabla}_{x_1} \widetilde{B}^{x_1 y_1} w^{x_2} d \Sigma_{x_2} \nonumber \\
&& + \int\limits_{\Sigma(s)} v^{y_2} \widehat{\nabla}_{y_2} \widetilde{B}^{x_1 y_1} d \Sigma_{x_1}.\label{int_aux}
\end{eqnarray}
Here $v^{y_1}:=dx^{y_1}/ds$, $s$ is the proper time, ${\frac{D}{ds}} = v^i\widehat{\nabla}_i$, and the integral is performed over a spatial hypersurface. Note that in our notation the point to which the index of a bitensor belongs can be directly read from the index itself; e.g., $y_{n}$ denotes indices at the point $y$. Furthermore, we will now associate the point $y$ with the world-line of the test body under consideration. 
Here the tilde marks densities, $\sigma$ denotes Synge's \cite{Synge:1960} world function, with $\sigma^y$ being its first covariant derivative, and $g^y{}_x$ is the parallel propagator for vectors. For objects with more complicated tensorial properties the parallel propagator is straightforwardly generalized to $G^Y{}_X$ and $G^{\dot{Y}}{}_{\dot{X}}$. We will need these generalized propagators to deal with the dynamical and material currents $J^{Aj}$ and $\Xi^{\dot{A}}$. More details are collected in appendix \ref{conventions_app}. 

After these preliminaries, we introduce integrated moments for the two types of currents via (for $n = 0,1,\dots)$
\begin{eqnarray}
j^{y_1\cdots y_n Y_0}  \!&=&\! (-1)^n\!\!\!\!\int\limits_{\Sigma(\tau)}\!\!\!\sigma^{y_1}\!\cdots\!\sigma^{y_n}G^{Y_0}{}_{X_0}\widetilde{J}^{X_0 x''}d\Sigma_{x''},\label{j1n}\\
i^{y_1\dots y_{n} Y_0 y'} \!&=&\! (-1)^n\!\!\!\!\int\limits_{\Sigma(\tau)}\!\!\!\sigma^{y_1}\!\cdots\!\sigma^{y_n}G^{Y_0}{}_{X_0}g^{y'}{}_{x'}\widetilde{J}^{X_0 x'}w^{x''}d\Sigma_{x''},\nonumber\\
&& \label{i1n}\\
m^{y_1\dots y_{n} \dot{Y}_0 } \!&=&\! (-1)^n\!\!\!\!\int\limits_{\Sigma(\tau)}\!\!\!\sigma^{y_1}\!\cdots\!\sigma^{y_n}G^{\dot{Y}_0}{}_{\dot{X}_0}\widetilde{\Xi}^{\dot{X}_0}w^{x''}d\Sigma_{x''}.\label{m1n}
\end{eqnarray}
Integrating (\ref{dJ}) and making use of (\ref{int_aux}), we find the following ``master equation of motion'' for the generalized multipole moments:
\begin{widetext}
\begin{eqnarray}
{\frac{D}{ds}} j^{y_1\cdots y_n Y_0}  \!&=&\! - n\, v^{(y_1} j^{y_2 \dots y_n) Y_0} + n\, i^{(y_1 \dots y_{n-1}|Y_0|y_n)}- \gamma^{Y_0}{}_{Y'y''y_{n+1}}\left(i^{y_1 \dots y_{n}Y'y''} + j^{y_1 \dots y_{n}Y'}v^{y''}\right)\nonumber\\
&&- \Lambda_{y'Y''}{}^{Y_0}i^{y_1 \dots y_{n}Y''y'} - \Lambda_{y'Y''}{}^{Y_0}{}_{;y_{n+1}}i^{y_1 \dots y_{n+1}Y''y'}- \Pi^{Y_0}{}_{\dot{Y}'}m^{y_1 \dots y_{n}\dot{Y}'} - \Pi^{Y_0}{}_{\dot{Y}';y_{n+1}}m^{y_1 \dots y_{n+1}\dot{Y}'}\nonumber\\ 
&&+ \sum\limits^{\infty}_{k=2}{\frac 1{k!}}\Bigl[-(-1)^k n\, \alpha^{(y_1}{}_{y' y_{n+1} \dots y_{n+k}} i^{y_2 \dots y_n)y_{n+1} \dots y_{n+k} Y_0 y'}+ (-1)^k n\, v^{y'} \beta^{(y_1}{}_{y' y_{n+1} \dots y_{n+k}} j^{y_2 \dots y_n)y_{n+1} \dots y_{n+k} Y_0}\nonumber\\ 
&&+ (-1)^k\gamma^{Y_0}{}_{Y'y''y_{n+1}\dots y_{n+k}}\left(i^{y_1 \dots y_{n+k}Y'y''} + j^{y_1 \dots y_{n+k}Y'}v^{y''}\right)- \Lambda_{y'Y''}{}^{Y_0}{}_{;y_{n+1}\dots y_{n+k}}i^{y_1 \dots y_{n+k}Y''y'}\nonumber\\ 
&&- \Pi^{Y_0}{}_{\dot{Y}';y_{n+1}\dots y_{n+k}}m^{y_1 \dots y_{n+k}\dot{Y}'}\Bigr].\label{master}
\end{eqnarray}
\end{widetext}

\subsection{Electrodynamics in Minkowski spacetime}\label{eom_max}

To see how the general formalism works, let us consider the motion of electrically charged extended bodies under the influence of electromagnetic field in the flat Minkowski spacetime. This problem was analyzed earlier by means of a different approach in \cite{Dixon:1967}. 

In this case, it is convenient to recast the set of dynamical currents into the form of a column
\begin{equation}
J^{Aj} = \left(\begin{array}{c}J^j \\ \Sigma^{kj}\end{array}\right),\label{Jmax}
\end{equation}
where $J^j$ is the electric current and $\Sigma^{kj}$ is the energy-momentum tensor. Physically, the structure of the dynamical current is crystal clear: the matter elements of an extended body are characterized by the two types of ``charges'', the electrical charge (the upper component) and the mass (the lower component). 

The generalized conservation law comprises two components of different tensor dimensions:
\begin{equation}
\widehat{\nabla}_j \left(\begin{array}{c}J^j \\ \Sigma^{kj}\end{array}\right) = 
\left(\begin{array}{c}0 \\ - F^{kj}J_j\end{array}\right),\label{dJmax}
\end{equation}
where the lower component of the right-hand side describes the usual Lorentz force. 

Accordingly, we indeed recover for the dynamical current (\ref{Jmax}) the conservation law in the form (\ref{dJ}) where $\Xi^{\dot{B}} = 0$ and
\begin{equation}
\Lambda_{jB}{}^A = \left(\begin{array}{c|c}0 & 0\\ \hline F_j{}^k & 0\end{array}\right).\label{LamMax}
\end{equation}
The generalized moments (\ref{j1n})-(\ref{m1n}) have the same column structure, reflecting the two physical charges of matter:
\begin{eqnarray}
j^{y_1\cdots y_n Y_0}  \!&=&\! \left(\begin{array}{c}j^{y_1\cdots y_n} \\ p^{y_1\cdots y_ny_0} \end{array}\right),\label{jMmax}\\
i^{y_1\cdots y_n Y_0y'}  \!&=&\! \left(\begin{array}{c}i^{y_1\cdots y_ny'} \\ k^{y_1\cdots y_ny_0y'} \end{array}\right),\label{iMmax}
\end{eqnarray}
whereas $m^{y_1\dots y_{n} \dot{Y}_0 } =0$.

As a result, the master equation (\ref{master}) reduces to the coupled system of the two
sets of equations for the moments:
\begin{eqnarray}
{\frac{D}{ds}} j^{y_1\cdots y_n}  &=& - n\, v^{(y_1} j^{y_2 \dots y_n)} + n\,i^{(y_1 \dots y_{n})},\label{dj1nmax}\\
{\frac{D}{ds}} p^{y_1\cdots y_n y_0}  &=& - n\, v^{(y_1} p^{y_2 \dots y_n)y_0} + n\,k^{(y_1 \dots y_{n-1}|y_0| y_{n})}\nonumber\\
&&  - \sum\limits^{\infty}_{k=1}{\frac 1{k!}}F_{y'}{}^{y_0}{}_{;y_{n+1}\dots y_{n+k}}i^{y_1 \dots y_{n+k}y'} \nonumber \\
&&- F_{y'}{}^{y_0}i^{y_1 \dots y_{n}y'}.\label{dp1nmax}
\end{eqnarray}
These equations should be compared to those of \cite{Dixon:1967}.

\section{Equations of motion in metric-affine gravity}\label{eom_sec}

We are now in a position to derive the equations of motion for extended test bodies in metric-affine gravity. Introducing the dynamical current
\begin{equation}
J^{Aj} = \left(\begin{array}{c}\Delta^{ikj} \\ \Sigma^{kj}\end{array}\right),\label{JAM}
\end{equation}
and the material current
\begin{equation}
\Xi^{\dot{A}} = \left(\begin{array}{c} t^{ik} \\ L_{\rm mat}\end{array}\right),\label{XIM}
\end{equation}
we then recast the system (\ref{cons1f}) and (\ref{cons2f}) into the generic conservation law (\ref{dJ}), where we now have 
\begin{eqnarray}
\Lambda_{jB}{}^A &=& \left(\begin{array}{c|c}U_{ji'k'}{}^{ik} & - \delta^i_j\delta^k_{k'}\\ \hline R^k{}_{ji'k'} & V_{jk'}{}^k\end{array}\right),\label{LamMAG}\\
  \Pi^A{}_{\dot{B}} &=& \left(\begin{array}{c|c}\delta^i_{i'}\delta^k_{k'} & 0\\ \hline {\frac 12} Q^k{}_{i'k'} & A^k\end{array}\right).\label{PIMAG}
\end{eqnarray}

Like in the previous example of an electrically charged body, the matter elements in metric-affine gravity are also characterized by two ``charges'': the canonical hypermomentum (upper component) and the canonical energy-momentum (lower component). This is reflected in the column structure of the dynamical current (\ref{JAM}). The material current (\ref{XIM}) takes into account the metrical energy-momentum and the matter Lagrangian related to the nonminimal coupling. The multi-index $A = \{ik,k\}$, whereas $\dot{A} = \{ik,1\}$. Accordingly, the generalized propagator reads
\begin{equation}
G^Y{}_X = \left(\begin{array}{c|c} g^{y_1}{}_{x_1}g^{y_2}{}_{x_2} & 0\\ \hline 0 & g^{y_1}{}_{x_1}\end{array}\right),\label{propMAG}
\end{equation}
and we easily construct the expansion coefficients of its derivatives from the corresponding expansions of the derivatives of the vector propagator $g^{y}{}_{x}$:
\begin{eqnarray}
\gamma^{Y_0}{}_{Y_1y_2\dots y_{k+2}} = \left(\begin{array}{c|c} \gamma^{\{y_0\tilde{y}\}}{}_{\{y'y''\}y_2\dots y_{k+2}} & 0\\ \hline 0 & \gamma^{y_0}{}_{y'y_2\dots y_{k+2}}\end{array}\right),\nonumber\\ \label{propE}
\end{eqnarray}
where we denoted 
\begin{equation}
\gamma^{\{y_0\tilde{y}\}}{}_{\{y'y''\}y_2\dots y_{k+2}} = \gamma^{y_0}{}_{y'y_2\dots y_{k+2}}\delta^{\tilde{y}}_{y''} +  \gamma^{\tilde{y}}{}_{y''y_2\dots y_{k+2}}\delta^{y_0}_{y'}.\label{ggg}
\end{equation}
In particular, for the first expansion coefficient ($k = 1$), we find
\begin{eqnarray}
\gamma^{\{y_0\tilde{y}\}}{}_{\{y'y''\}y_2y_{3}} &=& {\frac 12}\left(\hat{R}^{y_0}{}_{y'y_2y_3}\delta^{\tilde{y}}_{y''} + \hat{R}^{\tilde{y}}{}_{y''y_2y_3}\delta^{y_0}_{y'}\right),\label{gR1}\\
\gamma^{y_0}{}_{y'y_2y_{3}} &=& {\frac 12}\hat{R}^{y_0}{}_{y'y_2y_3}.\label{gR2}
\end{eqnarray}

For completeness, let us also write down another generalized propagator 
\begin{equation}
G^{\dot{Y}}{}_{\dot{X}} = \left(\begin{array}{c|c} g^{y_1}{}_{x_1}g^{y_2}{}_{x_2} & 0\\ \hline 0 & 1\end{array}\right).\label{propMAG2}
\end{equation}

The last step is to write the generalized moments (\ref{j1n})-(\ref{m1n}) in terms of their components:
\begin{eqnarray}
j^{y_1\cdots y_n Y}  \!&=&\! \left(\begin{array}{c}h^{y_1\cdots y_ny'y''} \\ p^{y_1\cdots y_ny'} \end{array}\right),\label{jMAG}\\
i^{y_1\dots y_{n} Y y_0} \!&=&\! \left(\begin{array}{c}q^{y_1\cdots y_ny'y''y_0} \\ k^{y_1\cdots y_ny'y_0} \end{array}\right),\label{iMAG}\\
m^{y_1\dots y_{n} \dot{Y} } \!&=&\! \left(\begin{array}{c}\mu^{y_1\cdots y_ny'y''} \\ \xi^{y_1\cdots y_n} \end{array}\right).\label{mMAG}
\end{eqnarray}
For the two most important moments, ``$h$'' stands for the hypermomentum, whereas ``$p$'' stands for the momentum.\footnote{Note that in order to facilitate the comparison with our previous work \cite{Puetzfeld:Obukhov:2013:3}, we provide in appendix \ref{explicit_app} the explicit form of integrated conservation laws (\ref{cons1f}) and (\ref{cons2f}), as well as the generalized integrated moments (\ref{jMAG}) -- (\ref{mMAG}) in the notation used in \cite{Puetzfeld:Obukhov:2013:3}.} Finally, substituting all of the above into the ``master equation'' (\ref{master}), we obtain the system of multipolar equations of motion for extended test bodies in metric-affine gravity:
\begin{widetext}
\begin{eqnarray}
\frac{D}{ds} h^{y_1 \dots y_n y_a y_b} &=&  - n \, v^{(y_1} h^{y_2 \dots y_n) y_a y_b} + n \, q^{(y_1 \dots y_{n-1} | y_a y_b | y_n)} + k^{y_1 \dots y_n y_b y_a} - \mu^{y_1 \dots y_n y_a y_b}  \nonumber \\
&& - \frac{1}{2}\widehat{R}^{y_a}{}_{y' y'' y_{n+1}} \left(q^{y_1 \dots y_{n+1} y' y_b y''} + v^{y''} h^{y_1 \dots y_{n+1} y' y_b}\right)  \nonumber \\
&& - \frac{1}{2}\widehat{R}^{y_b}{}_{y' y'' y_{n+1}}\left( q^{y_1 \dots y_{n+1} y_a y' y''} + v^{y''} h^{y_1 \dots y_{n+1} y_a y'}\right)   \nonumber \\
&& - U_{y_0y'y''}{}^{y_ay_b} q^{y_1 \dots y_n y' y'' y_0} - U_{y_0y'y''}{}^{y_ay_b}{}_{;y_{n+1}} q^{y_1 \dots y_{n+1} y' y'' y_0}\nonumber\\
&& + \sum^{\infty}_{k=2} {\frac{1}{k!}}\Bigg[  (-1)^k \gamma^{y_a}{}_{y' y'' y_{n+1} \dots y_{n+k}}\left( q^{y_1 \dots y_{n+k} y' y_b y''} + v^{y''} h^{y_1 \dots y_{n+k} y' y_b}\right) \nonumber \\
&& + (-1)^k \gamma^{y_b}{}_{y' y'' y_{n+1} \dots y_{n+k}}\left( q^{y_1 \dots y_{n+k} y_a y' y''} + v^{y''} h^{y_1 \dots y_{n+k} y_a y'}\right)   \nonumber \\
&& - (-1)^k n \, \alpha^{(y_1}{}_{y' y_{n+1} \dots y_{n+k}} q^{y_2 \dots y_n)y_{n+1} \dots y_{n+k} y_a y_b y'} + (-1)^k  v^{y'} n \, \beta^{(y_1}{}_{y' y_{n+1} \dots y_{n+k}} h^{y_2 \dots y_n)y_{n+1} \dots y_{n+k} y_a y_b} \nonumber \\
&& -  U_{y_0y'y''}{}^{y_ay_b}{}_{;y_{n+1} \dots y_{n+k}} q^{y_1 \dots y_{n+k} y' y'' y_0} \Bigg], \label{int_eom_1_general}
\end{eqnarray}
\begin{eqnarray}
\frac{D}{ds} p^{y_1 \dots y_n y_a}&=& - n \, v^{(y_1} p^{y_2 \dots y_n) y_a} + n \, k^{(y_1 \dots y_{n-1} | y_a | y_n)} - A^{y_a} \xi^{y_1 \dots y_n} - A^{y_a}{}_{;y_{n+1}} \xi^{y_1 \dots y_{n+1}} \nonumber\\
&&  - V_{y''y'}{}^{y_a} k^{y_1 \dots y_n y' y''} - V_{y''y'}{}^{y_a}{}_{;y_{n+1}} k^{y_1 \dots y_{n+1} y' y''} - \frac{1}{2} \widehat{R}^{y_a}{}_{y' y'' y_{n+1}}\left(k^{y_1 \dots y_{n+1} y' y''} + v^{y''} p^{y_1 \dots y_{n+1} y'} \right)\nonumber \\
&&   - R^{y_a}{}_{y_0 y' y''} q^{y_1 \dots y_n y' y'' y_0} - R^{y_a}{}_{y_0 y' y'';y_{n+1}} q^{y_1 \dots y_{n+1} y' y'' y_0} \nonumber \\
&&  - \frac{1}{2} Q^{y_a}{}_{y'' y'} \mu^{y_1 \dots y_n y' y''}  - \frac{1}{2} Q^{y_a}{}_{y'' y';y_{n+1}} \mu^{y_1 \dots y_{n+1} y' y''} \nonumber \\
&& + \sum^{\infty}_{k=2} \frac{1}{k!}\Bigg[ (-1)^k \gamma^{y_a}{}_{y' y'' y_{n+1} \dots y_{n+k}}\left( k^{y_1 \dots y_{n+k} y' y''} +  v^{y''} p^{y_1 \dots y_{n+k} y'}\right) \nonumber \\ 
&&  - (-1)^k n \, \alpha^{(y_1}{}_{y' y_{n+1} \dots y_{n+k}} k^{y_2 \dots y_n)y_{n+1} \dots y_{n+k} y_a y' } + (-1)^k n \, v^{y'} \beta^{(y_1}{}_{y' y_{n+1} \dots y_{n+k}} p^{y_2 \dots y_n ) y_{n+1} \dots y_{n+k} y_a} \nonumber \\
&&  - R^{y_a}{}_{y_0 y' y'';y_{n+1} \dots y_{n+k}} q^{y_1 \dots y_{n+k} y' y'' y_0} - V_{y''y'}{}^{y_a}{}_{;y_{n+1} \dots y_{n+k}} k^{y_1 \dots y_{n+k} y' y''} \nonumber \\
&&  - \frac{1}{2} Q^{y_a}{}_{y'' y';y_{n+1} \dots y_{n+k}} \mu^{y_1 \dots y_{n+k} y' y''}  - A^{y_a}{}_{;y_{n+1} \dots y_{n+k}} \xi^{y_1 \dots y_{n+k}}  \Bigg]. 
\label{int_eom_2_general}
\end{eqnarray}
\end{widetext}

\section{Special cases} \label{special_cases_sec}

The general equations of motion (\ref{int_eom_1_general}) and (\ref{int_eom_2_general}) are valid to {\it any} multipolar order. In the following sections we focus on some special cases; in particular, we work out the two lowest multipolar orders of approximation and consider the explicit form of the equations of motion in special geometries.

\subsection{General pole-dipole equations of motion}

From (\ref{int_eom_1_general}) and (\ref{int_eom_2_general}), we can derive the general pole-dipole equations of motion. The relevant moments to be kept at this order of approximation are $p^a, p^{ab}, h^{ab}, q^{abc}, k^{ab}, k^{abc}, \mu^{ab}, \mu^{abc}, \xi^{a},$ and $\xi$. Since all objects are now evaluated on the world-line, we switch back to the usual tensor notation.

For $n=1$ and $n=0$, eq.\ (\ref{int_eom_1_general}) yields
\begin{eqnarray}
0 &=& k^{acb} - \mu^{abc} + q^{bca} - v^a h^{bc}, \label{eom_1_n_1} \\
\frac{D}{ds} h^{ab} &=&  k^{ba} - \mu^{ab} - U_{cde}{}^{ab} q^{dec}.  \label{eom_1_n_0}
\end{eqnarray}
Furthermore for $n=2,1,0$ equation (\ref{int_eom_2_general}) yields
\begin{eqnarray}
0 &=& k^{(a|c|b)} - v^{(a} p^{b)c}, \label{eom_2_n_2} \\
\frac{D}{ds} p^{ab} &=&  k^{ba} - v^a p^b  - A^b \xi^a  - V_{dc}{}^b k^{acd} - \frac{1}{2} Q^b{}_{dc} \mu^{acd},\nonumber \\
\label{eom_2_n_1} \\
\frac{D}{ds} p^{a} &=& - V_{cb}{}^a k^{bc} - R^{a}{}_{dbc} q^{bcd} - \frac{1}{2} Q^a{}_{cb} \mu^{bc} \nonumber \\
&& - A^a \xi -\frac{1}{2} \widehat{R}^a{}_{cdb} \left(k^{bcd} + v^d p^{bc} \right) \nonumber \\
&&- V_{dc}{}^a{}_{;b} k^{bcd} - \frac{1}{2} Q^a{}_{dc;b} \mu^{bcd} - A^a{}_{;b} \xi^b. \label{eom_2_n_0} 
\end{eqnarray}

\subsubsection{Rewriting equations of motion}

Let us decompose (\ref{eom_1_n_1}) and (\ref{eom_1_n_0}) into symmetric and skew-symmetric parts:
\begin{eqnarray}
\mu^{abc} &=& k^{a(bc)} + q^{(bc)a} - v^a h^{(bc)}, \label{eom_1_n_1S} \\
0 &=& -\,k^{a[bc]} + q^{[bc]a} - v^a h^{[bc]}, \label{eom_1_n_1A} \\
\mu^{ab} &=& -\,\frac{D}{ds}h^{(ab)} +  k^{(ab)} - U_{cde}{}^{(ab)} q^{dec},  \label{eom_1_n_0S}\\
\frac{D}{ds} h^{[ab]} &=& -\,k^{[ab]} - U_{cde}{}^{[ab]} q^{dec}.  \label{eom_1_n_0A}
\end{eqnarray}
As a result, we can express the moments symmetric in the last two indices $\mu^{ab} = \mu^{(ab)}$ and $\mu^{cab} = \mu^{c(ab)}$ (in general, this is possible also for an arbitrary order $\mu^{c_1\dots c_nab} = \mu^{c_1\dots c_n(ab)}$) in terms of the other moments. 

Let us denote the skew-symmetric part $s^{ab} := h^{[ab]}$, as this greatly simplifies the subsequent manipulations and the comparison with  \cite{Puetzfeld:Obukhov:2013:3}.

The system of the two equations (\ref{eom_2_n_2}) and (\ref{eom_1_n_1A}) can be resolved in terms of the 3rd rank $k$-moment. The result reads explicitly
\begin{eqnarray}
k^{abc} &=& v^a p^{cb} + v^c\left(p^{[ab]} - s^{ab}\right)\nonumber\\
&& + v^b\left(p^{[ac]} - s^{ac}\right) + v^a\left(p^{[bc]} - s^{bc}\right)\nonumber\\
&& + q^{[ab]c} + q^{[ac]b} + q^{[bc]a}.\label{kabc}
\end{eqnarray}
This yields some useful relations:
\begin{eqnarray}
k^{a[bc]} &=& - v^as^{bc} + q^{[bc]a},\label{ka1}\\
k^{[ab]c} &=&  v^{[a} p^{|c|b]} + v^c\left(p^{[ab]} - s^{ab}\right) + q^{[ab]c}.\label{ka2}
\end{eqnarray}

The next step is to use the equations (\ref{eom_1_n_0S}), (\ref{eom_1_n_1S}) together with (\ref{kabc}) and substitute the $\mu$-moments and $k$-moments into (\ref{eom_1_n_0}) and (\ref{eom_2_n_1})-(\ref{eom_2_n_0}). This yields the system that depends only on the $p,h,q$ and $\xi$ moments. 

Let us start with the analysis of (\ref{eom_2_n_0}). The latter contains the combination $k^{[b|c|d]} + v^{[d}p^{b]c}$ where the skew symmetry is imposed by the contraction with the Riemann curvature tensor which is antisymmetric in the last two indices. Making use of (\ref{kabc}), we derive
\begin{equation}
k^{[a|c|b]} + v^{[b}p^{a]c} = \kappa^{abc} + \kappa^{acb} - \kappa^{bca},\label{kka}
\end{equation}
where we introduced the abbreviation
\begin{equation}
\kappa^{abc} = v^c\left(p^{[ab]} - s^{ab}\right) +  q^{[ab]c}.\label{kappa}
\end{equation}
Note that by construction $\kappa^{abc} = \kappa^{[ab]c}$. 

Then by making use of the Ricci identity we find 
\begin{eqnarray}
-\,{\frac 12} \widehat{R}^a{}_{cdb} \left(k^{bcd} + v^d p^{bc} \right) &=& 
\widehat{R}^a{}_{bcd}\left[q^{[cd]b} \right.\nonumber\\
&&\left. + v^b\left(p^{[cd]} - s^{cd}\right)\right].\label{Rq}
\end{eqnarray}

Substituting $k^{bc}$ from (\ref{eom_2_n_1}) and $\mu^{bc}$ from (\ref{eom_1_n_0S}), we find after some algebra
\begin{eqnarray}
&&-\,V_{cb}{}^ak^{bc} - {\frac 12}Q^a{}_{cb}\mu^{bc} = - \,A_b{\frac {Dp^{ba}}{ds}} - N^a{}_{cd}{\frac {Dh^{cd}}{ds}}\nonumber\\
&&- \left(p^a + N^a{}_{cd}h^{cd}\right)v^bA_b - A^aA^b\xi_b - k^{bac}A_bA_c\nonumber\\
&&+ \left(N^a{}_{nb}N_{dc}{}^n - N^a{}_{cn}N_d{}^n{}_b\right)q^{cbd}.\label{VkQmu1}
\end{eqnarray}
Further simplification is achieved by noticing that
\begin{eqnarray}
v^bA_b &=& {\frac {DA}{ds}},\label{vA}\\
k^{bac}A_bA_c &=& p^{ca}A_c{\frac {DA}{ds}},\label{kAA}
\end{eqnarray}
where we used (\ref{eom_2_n_2}) and recalled that $A_b = A_{;b}$.

Analogously, taking $k^{b[cd]}$ from (\ref{ka1}) and $\mu^{bcd}$ from (\ref{eom_1_n_1S}), we derive
\begin{eqnarray}
&&-\,V_{dc}{}^a{}_{;b}k^{bcd} - {\frac 12}Q^a{}_{dc;b}\mu^{bcd} = - \,A_{b;c}k^{cab} \nonumber\\
&& + N^a{}_{cd;b}q^{cdb} - N^a{}_{cd;b}v^bh^{cd}.\label{VkQmu2}
\end{eqnarray}
We can again use $A_b = A_{;b}$ and (\ref{eom_2_n_2}) to simplify
\begin{equation}
- \,A_{b;c}k^{cab} = - p^{ba}{\frac {DA_b}{ds}}.\label{kAd}
\end{equation}

After these preliminary calculations, we substitute (\ref{Rq})-(\ref{kAd}) into (\ref{eom_2_n_0}) to recast the latter into 
\begin{eqnarray}
&& {\frac {D}{ds}}\left(Fp^a + FN^a{}_{cd}h^{cd} + p^{ba}\widehat{\nabla}_bF\right)  \nonumber\\
&& = F \widehat{R}^a{}_{bcd}v^b\left(p^{[cd]} - s^{cd}\right)\nonumber\\
&& - Fq^{cbd}\left[R^a{}_{dcb} - \widehat{R}^a{}_{dcb} - N^a{}_{cb;d}\right.\nonumber\\
&& \left. - N^a{}_{nb}N_{dc}{}^n + N^a{}_{cn}N_d{}^n{}_b\right]\nonumber\\
&& - FA^a\left(\xi + \xi^bA_b\right) - F\xi^bA^a{}_{;b}.\label{DPa}
\end{eqnarray}

Finally, combining (\ref{eom_2_n_1}) and (\ref{eom_1_n_0}) to eliminate $k^{ba}$ we derive the equation
\begin{eqnarray}
{\frac {D}{ds}}\left(p^{ab} - h^{ab}\right) &=& \mu^{ab} - v^a\left(p^b + N^b{}_{cd}h^{cd}\right)\nonumber\\
&& +\,q^{cda}N^b{}_{cd} - q^{cbd}N_{dc}{}^a + q^{acd}N_d{}^b{}_c\nonumber\\
&& -\,\xi^aA^b + (q^{abc} - k^{abc})A_c.\label{Dpab}
\end{eqnarray}

Following \cite{Puetzfeld:Obukhov:2013:3}, we introduce the total orbital and the total spin angular moments
\begin{equation}
L^{ab} := 2p^{[ab]},\qquad S^{ab} := -2h^{[ab]},\label{LS}
\end{equation}
and define the generalized total energy-momentum 4-vector and the generalized total angular momentum by
\begin{eqnarray}
{\cal P}^a &:=& F(p^a + N^a{}_{cd}h^{cd}) + p^{ba}\widehat{\nabla}_bF,\label{Ptot}\\
{\cal J}^{ab} &:=& F(L^{ab} + S^{ab}).\label{Jtot}
\end{eqnarray}
Then, taking into account the identity (\ref{RRN}) which with the help of the raising and lowering of indices can be recast into
\begin{eqnarray}
\widehat{\nabla}^aN_{dcb} &=& - R^a{}_{dcb} + \widehat{R}^a{}_{dcb} + N^a{}_{cb;d}\nonumber\\
&& + N^a{}_{nb}N_{dc}{}^n - N_d{}^n{}_bN^a{}_{cn},\label{DNRR}
\end{eqnarray}
we rewrite the pole-dipole equations of motion (\ref{DPa}) and (\ref{Dpab}) in the final form
\begin{eqnarray}
{\frac {D{\cal P}^a}{ds}} &=& {\frac 12}\widehat{R}^a{}_{bcd}v^b{\cal J}^{cd} + Fq^{cbd}\widehat{\nabla}^aN_{dcb} \nonumber\\ 
&& -\, \xi\widehat{\nabla}^aF - \xi^b\widehat{\nabla}_b\widehat{\nabla}^aF,\label{DPtot}\\
{\frac {D{\cal J}^{ab}}{ds}} &=& -\,2v^{[a}{\cal P}^{b]} + 2F(q^{cd[a}N^{b]}{}_{cd} + q^{c[a|d|}N_{dc}{}^{b]}\nonumber\\
&& + q^{[a|cd|}N_d{}^{b]}{}_c) - 2\xi^{[a}\widehat{\nabla}^{b]}F.\label{DJtot}
\end{eqnarray}
The last equation arises as the skew-symmetric part of (\ref{Dpab}), whereas the symmetric part of the latter is a non-dynamical relation that determines the $\mu^{ab}$ moment
\begin{eqnarray}
\mu^{ab} &=& {\frac {D\Upsilon^{ab}}{ds}} + {\frac 1F} v^{(a} \left( {\cal P}^{b)} + {\cal J}^{b)c}A_c \right) + \xi^{(a}A^{b)}\nonumber\\
&& -\,q^{cd(a}N^{b)}{}_{cd} + q^{c(a|d|}N_{dc}{}^{b)} - q^{(a|cd|}N_d{}^{b)}{}_c\nonumber\\ 
&& +\,(q^{[ac]b} + q^{[bc]a} - q^{(ab)c})A_c.\label{muabPD}
\end{eqnarray}
Here the symmetric moment of the total hypermomentum is introduced via
\begin{equation}
\Upsilon^{ab} := p^{(ab)} - h^{(ab)}.\label{hypermom}
\end{equation}

\subsection{Coupling to the post-Riemannian geometry: Fine structure}

Let us look more carefully at how the post-Riemannian pieces of the gravitational field couple to extended test bodies. At first, we notice that the generalized energy-momentum vector (\ref{Ptot}) contains the term $N^a{}_{cd}h^{cd}$ that describes the direct interaction of the distortion (torsion plus nonmetricity) with the intrinsic dipole moment of the hypermomentum. Decomposing the latter into the skew-symmetric (spin) part and the symmetric (proper hypermomentum + dilation) part, we find
\begin{equation}
N^a{}_{cd}h^{cd} = -\,{\frac 12}N^a{}_{[cd]}S^{cd} - {\frac 12}Q^a{}_{cd}h^{(cd)}.\label{Nh}
\end{equation}
Here we made use of (\ref{QN}). This is quite consistent with the gauge-theoretic structure of metric-affine gravity. The second term shows that the intrinsic proper hypermomentum and the dilation moment couple to the nonmetricity, whereas the first term displays the typical spin-torsion coupling. 

Similar observations can be made for the coupling of higher moments which appear on the right-hand sides of (\ref{DPtot}) and (\ref{DJtot}) - and thus determine the force and torque acting on an extended body due to the post-Riemannian gravitational field. In order to see this, let us introduce the decomposition 
\begin{equation}
-\,{\frac 12}q^{abc} = {\stackrel {d}{q}}{}^{abc} + {\stackrel {s}{q}}{}^{cab}\label{qqq}
\end{equation}
into the two pieces 
\begin{eqnarray}
{\stackrel {d}{q}}{}^{abc} &:=& {\frac 12}\left(q^{[ac]b} + q^{[bc]a} - q^{(ab)c}\right),\label{qd}\\
{\stackrel {s}{q}}{}^{abc} &:=& {\frac 12}\left(q^{[ab]c} + q^{[ac]b} - q^{[bc]a}\right).\label{qs}
\end{eqnarray}
The overscript ``$d$'' and ``$s$'' notation shows the relevance of these objects to the dilation plus proper hypermomentum and to the spin, respectively. By construction, we have the following algebraic properties
\begin{equation}
{\stackrel {d}{q}}{}^{[ab]c} \equiv 0,\qquad {\stackrel {s}{q}}{}^{(ab)c}  \equiv 0.\label{qq0}
\end{equation}

Making use of the decomposition (\ref{qqq}) and of the explicit structure of the distortion (\ref{NTQ}), we then recast the equations of motion (\ref{DPtot}) and (\ref{DJtot}) into
\begin{eqnarray}
{\frac {D{\cal P}^a}{ds}} &=& {\frac 12}\widehat{R}^a{}_{bcd}v^b{\cal J}^{cd}\nonumber\\ 
&& +\,F{\stackrel {s}{q}}{}^{cbd}\widehat{\nabla}^a T_{cbd} + F{\stackrel {d}{q}}{}^{cbd}\widehat{\nabla}^a Q_{dcb}\nonumber\\ 
&& -\,\xi\widehat{\nabla}^aF - \xi^b\widehat{\nabla}_b\widehat{\nabla}^aF,\label{DPdec}\\
{\frac {D{\cal J}^{ab}}{ds}} &=& -\,2v^{[a}{\cal P}^{b]}\nonumber\\ 
&& +\,2F({\stackrel {s}{q}}{}^{cd[a}T_{cd}{}^{b]} + 2{\stackrel {s}{q}}{}^{[a|cd|}T^{b]}{}_{cd})\nonumber\\
&& +\, 2F({\stackrel {d}{q}}{}^{cd[a}Q^{b]}{}_{cd} + 2{\stackrel {d}{q}}{}^{[a|dc|}Q_{cd}{}^{b]})\nonumber\\ 
&& -\,2\xi^{[a}\widehat{\nabla}^{b]}F.\label{DJdec}
\end{eqnarray}
Now we clearly see the fine structure of the coupling of extended bodies to the post-Riemannian geometry. The first lines in the equations of motion describe the usual Mathisson-Papapetrou force and torque. They depend on the Riemannian geometry only. A body with the nontrivial moment (\ref{qs}) is affected by the torsion field, whereas the nontrivial moment (\ref{qd}) feels the nonmetricity. This explains the different physical meaning of the higher moments (\ref{qd}) and (\ref{qs}). In addition, the last lines in (\ref{DPdec}) and (\ref{DJdec}) describe contributions due to the nonminimal coupling. 

\subsection{General monopolar equations of motion}

At the monopolar order we have nontrivial moments $p^a, k^{ab}, \mu^{ab}$ and $\xi$. The nontrivial equations of motion then arise from the eq.\ (\ref{int_eom_1_general}) for $n = 0$ and from the eq.\ (\ref{int_eom_2_general}) for $n=1, n = 0$:
\begin{eqnarray}
0 &=& k^{ba} - \mu^{ab},\label{Mono1}\\
0 &=& k^{ba} - v^a p^b,\label{Mono2}\\
{\frac {Dp^a}{ds}} &=& -\,V_{cb}{}^ak^{bc} - {\frac 12}Q^a{}_{cb}\mu^{bc} - A^a\xi.\label{Mono3}
\end{eqnarray}
The first two equations (\ref{Mono1}) and (\ref{Mono2}) yield
\begin{equation}
k^{[ab]} = 0,\qquad v^{[a} p^{b]} = 0,\label{vp}
\end{equation}
and substituting (\ref{Mono1}), (\ref{Mono2}) and (\ref{vp}) into (\ref{Mono3}) we find
\begin{equation}
{\frac {D(Fp^a)}{ds}} = -\,\xi\widehat{\nabla}^aF.\label{Mono4}
\end{equation}
From (\ref{vp}) we have $p^a = Mv^a$ with the mass $M := v^ap_a$, and this allows us to recast (\ref{Mono4}) into the final form
\begin{equation}
M{\frac {Dv^a}{ds}} = -\,\xi(g^{ab} - v^av^b){\frac {\widehat{\nabla}_bF}F}.\label{Mono5}
\end{equation}
Hence, in general the motion of nonminimally coupled monopole test bodies is nongeodetic. Furthermore, the general monopole equation of motion (\ref{Mono5}) reveals an interesting feature of theories with nonminimal coupling. There is an ``indirect'' coupling, i.e.\ through the coupling function $F(g_{ij}, R_{ijk}{}^l,T_{ij}{}^k,Q_{kij})$, of post-Riemannian spacetime features to structureless test bodies.  

\subsection{Weyl-Cartan spacetime}

In Weyl-Cartan spacetime the nonmetricity reads $Q_{kij} = Q_kg_{ij}$, where $Q_k$ is the Weyl covector. Hence the distortion is given by
\begin{eqnarray}\label{wc_distortion}
N_{kj}{}^i = K_{kj}{}^i + {\frac{1}{2}}\left(Q^i g_{kj} - Q_k\delta^i_j - Q_j\delta^i_k\right).
\end{eqnarray}
The contortion tensor is constructed from the torsion,
\begin{equation}
K_{kj}{}^i = -\,{\frac 12}(T_{kj}{}^i + T^i{}_{kj} + T^i{}_{jk}).\label{KT}
\end{equation}
As a result, the generalized momentum (\ref{Ptot}) in Weyl-Cartan spacetime takes the form
\begin{eqnarray}
{\cal P}^a = Fp^a - {\frac F2}\left(K^a{}_{cd} S^{cd} - Q_bS^{ba} + Q^aD\right) + p^{ba}\widehat{\nabla}_bF.\nonumber\\
 \label{wc_ptot}
\end{eqnarray}
Here we introduced the {\it intrinsic dilation} moment $D := g_{ab}h^{ab}$. 

Substituting the distortion (\ref{wc_distortion}) into (\ref{DPtot}) and (\ref{DJtot}), we find the pole-dipole equations of motion in the Weyl-Cartan spacetime:
\begin{eqnarray}
{\frac {D{\cal P}^a}{ds}} &=& {\frac 12}\widehat{R}^a{}_{bcd}v^b{\cal J}^{cd} + F{\stackrel {s}{q}}{}^{cbd}\widehat{\nabla}^a T_{cbd} \nonumber\\ 
&& +\,Z^b\widehat{\nabla}^aQ_b -\xi\widehat{\nabla}^aF - \xi^b\widehat{\nabla}_b\widehat{\nabla}^aF,\label{DPWC}\\
{\frac {D{\cal J}^{ab}}{ds}} &=& -\,2v^{[a}{\cal P}^{b]} + 2F({\stackrel {s}{q}}{}^{cd[a}T_{cd}{}^{b]} + 2{\stackrel {s}{q}}{}^{[a|cd|}T^{b]}{}_{cd})\nonumber\\
&& + 2FZ^{[a}Q^{b]} - 2\xi^{[a}\widehat{\nabla}^{b]}F.\label{DJWC}
\end{eqnarray}
Here we introduced the trace of the modified moment (\ref{qd}),
\begin{eqnarray}
Z^a := g_{bc}{\stackrel {d}{q}}{}^{bca} = {\frac 12}g_{bc}\left(q^{bac} - q^{bca} - q^{abc}\right).\label{Za}
\end{eqnarray}
It is coupled to the Weyl nonmetricity.

\subsection{Weyl spacetime}

Weyl spacetime \cite{Weyl:1923} is obtained as a special case of the results above for vanishing torsion. Hence the contortion is trivial, 
\begin{eqnarray}
K_{abc} = 0. \label{contortion}
\end{eqnarray}
Taking this into account, the generalized momentum (\ref{wc_ptot}) and the equations of motion (\ref{DPWC}) and (\ref{DJWC}) are simplified even further.

It is interesting to note that besides a direct coupling of the dilation moment to the Weyl nonmetricity on the right-hand sides of (\ref{DPWC}) and (\ref{DJWC}), there is also a nontrivial coupling of the spin to the nonmetricity in (\ref{wc_ptot}). 

\subsection{Riemann-Cartan spacetime}

Another special case is obtained when the Weyl vector vanishes $Q_a = 0$. Equations (\ref{wc_ptot})-(\ref{DJWC}) then reproduce {\it in a covariant way} the findings of Yasskin and Stoeger \cite{Stoeger:Yasskin:1980} when the coupling is minimal ($F=1$). For nonminimal coupling we recover our earlier results in \cite{Puetzfeld:Obukhov:2013:3}. 

\section{Conclusions}\label{conclusion_sec}

We have worked out covariant test body equations of motion for standard metric-affine gravity, as well as its extensions with nonminimal coupling. Our results cover a very large class of gravitational theories, and one can use them as a theoretical basis for systematic tests of gravity by means of extended deformable bodies. 

Furthermore, our work generalizes a whole set of works \cite{Bailey:Israel:1975,Stoeger:Yasskin:1979,Stoeger:Yasskin:1980,Puetzfeld:2007,Puetzfeld:Obukhov:2008:1,Puetzfeld:Obukhov:2008:2,Puetzfeld:Obukhov:2013:1,Hehl:Obukhov:Puetzfeld:2013,Puetzfeld:Obukhov:2013:3,Puetzfeld:Obukhov:2013:4,Puetzfeld:Obukhov:2014:1}.
In particular it can be viewed as a completion of the program initiated in \cite{Puetzfeld:2007}, in which a noncovariant Papapetrou-type \cite{Papapetrou:1951:3} approach was used. The general equations of motion (\ref{int_eom_1_general}) and (\ref{int_eom_2_general}) cover all of the previously reported cases. As demonstrated explicitly, the master equation (\ref{master}) allows for a quick adoption to any physical theory, as soon as the conservation laws and (multi-)current structure are fixed.

It is satisfying to see that in the context of nonminimal metric-affine gravity, one is able to recover the same indirect coupling -- as previously reported in \cite{Puetzfeld:Obukhov:2013:3} in the case of torsion -- of new geometrical quantities to regular matter via the coupling function $F$. This may be exploited to devise new strategies to detect post-Riemannian spacetime features in future experiments. We hope that our covariant unified framework sheds more light on the systematic test of theories which exhibit nonminimal coupling.

\section*{Acknowledgements}
This work was supported by the Deutsche Forschungsgemeinschaft (DFG) through the grant LA-905/8-1/2 (D.P.). 

\appendix

\section{Conventions \& Symbols}\label{conventions_app}

\begin{table}
\caption{\label{tab_symbols}Directory of symbols.}
\begin{ruledtabular}
\begin{tabular}{ll}
Symbol & Explanation\\
\hline
&\\
\hline
\multicolumn{2}{l}{{Geometrical quantities}}\\
\hline
$g_{a b}$ & Metric\\
$\sqrt{-g}$ & Determinant of the metric \\
$\delta^a_b$ & Kronecker symbol \\
$x^{a}$, $s$ & Coordinates, proper time \\
$\Gamma_{a b}{}^c$ & Connection \\
$N_{a b}{}^c$ & Distortion \\
$Q_{a b c}$ & Nonmetricity \\
$T_{a b}{}^c$ & Torsion \\
$R_{a b c}{}^d$& Curvature \\
$\sigma$ & World function\\
$H^{abc}{}_d, H^{ab}{}_c, M^{abc}, E^{ab}{}_c$ & Field momenta \\
$g^{y_0}{}_{x_0}$, $G^Y{}_X$ & Parallel propagator\\
$(\sigma^A{}_B)_i{}^j$ & Generators coord.\ transf.\\
\hline
\multicolumn{2}{l}{{Matter quantities}}\\
\hline
$\psi^A$ & General matter field \\
$\Sigma_a{}^b$, & Canonical energy-momentum \\
$\Delta^a{}_b{}^c$ & Canonical hypermomentum \\
$t_a{}^b$ & Metrical energy-momentum \\
$\Delta^a{}_b{}^c$ & Hypermomentum \\
$D$ & Intrinsic dilation moment\\
$P_a$ &  Momentum density \\
$J_a{}^b$ &  Hypermomentum density \\
$v^a$ & Velocity \\
$\Xi^A$ & Material currents \\
$\mathcal{P}^a$ & Gen.\ momentum \\
$\mathcal{J}^{ab}$ & Gen.\ total angular momentum \\
$\Upsilon^{ab}$ & Total hypermomentum moment \\
$M$ & Mass \\
$L$ & Lagrangian \\ 
$j^{\dots}, i^{\dots}, m^{\dots}, p^{\dots}, k^{\dots}, $& Integrated moments\\
$h^{\dots}, q^{\dots}, \mu^{\dots}, \xi^{\dots}$  & \\
\hline
\multicolumn{2}{l}{{Auxiliary quantities}}\\
\hline
${\stackrel * \nabla}{}_a$ & Modified cov. derivative \\
$F$, $A$ & Coupling function\\
$J^{Aj}$ & Dynamical currents \\
$\alpha^{y_0}{}_{y_1 \dots y_n}$, $\beta^{y_0}{}_{y_1 \dots y_n}$, $\gamma^{y_0}{}_{y_1 \dots y_n}$& Expansion coefficients\\
$U_{abc}{}^d$, $V_{ab}{}^c$, $\Lambda_{jB}{}^A$, $\Pi^A{}_{B}$ & Auxiliary variables \\
$\Phi^{y_1 \dots y_n y_ 0}{}_{x_0}$, $\Psi^{y_2 \dots y_n+1 y_ 0 y_1}{}_{x_0 x_1}$, & \\
${\stackrel {d}{q}}{}^{abc}$, ${\stackrel {s}{q}}{}^{abc}$, $Z^a$ & \\
\hline
\multicolumn{2}{l}{{Operators}}\\
\hline
$\partial_i$, $\nabla_i$ & (Partial, covariant) derivative \\ 
$\frac{D}{ds} = $``$\dot{\phantom{a}}$'' & Total derivative \\
``$[ \dots ]$''& Coincidence limit\\
``$\widehat{\phantom{AA}}$'' & Riemannian quantity\\
``$\widetilde{\phantom{AA}}$'' & Density
\end{tabular}
\end{ruledtabular}
\end{table}
 
In the following we summarize our conventions, and collect some frequently used formulas. A directory of symbols used throughout the text can be found in table \ref{tab_symbols}.

For an arbitrary $k$-tensor $T_{a_1 \dots a_k}$, the symmetrization and antisymmetrization are defined by
\begin{eqnarray}
T_{(a_1\dots a_k)} &:=& {\frac 1{k!}}\sum_{I=1}^{k!}T_{\pi_I\!\{a_1\dots a_k\}},\label{S}\\
T_{[a_1\dots a_k]} &:=& {\frac 1{k!}}\sum_{I=1}^{k!}(-1)^{|\pi_I|}T_{\pi_I\!\{a_1\dots a_k\}},\label{A}
\end{eqnarray}
where the sum is taken over all possible permutations (symbolically denoted by $\pi_I\!\{a_1\dots a_k\}$) of its $k$ indices. As is well known, the number of such permutations is equal to $k!$. The sign factor depends on whether a permutation is even ($|\pi| = 0$) or odd ($|\pi| = 1$). The number of independent components of the totally symmetric tensor $T_{(a_1\dots a_k)}$ of rank $k$ in $n$ dimensions is equal to the binomial coefficient ${{n-1+k}\choose{k}} = (n-1+k)!/[k!(n-1)!]$, whereas the number of independent components of the totally antisymmetric tensor $T_{[a_1\dots a_k]}$ of rank $k$ in $n$ dimensions is equal to the binomial coefficient ${{n}\choose{k}} = n!/[k!(n-k)!]$. For example, for a second rank tensor $T_{ab}$ the symmetrization yields a tensor $T_{(ab)} = {\frac 12}(T_{ab} + T_{ba})$ with 10 independent components, and the antisymmetrization yields another tensor $T_{[ab]} = {\frac 12}(T_{ab} - T_{ba})$ with 6 independent components.
 
The covariant derivative defined by the Riemannian connection is conventionally denoted by the nabla or by the semicolon: $\widehat{\nabla}_a =$ ``$ {}_{;a}$''.

Our conventions for the Riemann curvature are as follows:
\begin{eqnarray}
&& 2 A^{c_1 \dots c_k}{}_{d_1 \dots d_l ; [ba] } \equiv 2 \widehat{\nabla}_{[a} \widehat{\nabla}_{b]} A^{c_1 \dots c_k}{}_{d_1 \dots d_l} \nonumber \\
& = & \sum^{k}_{i=1} \widehat{R}_{abe}{}^{c_i} A^{c_1 \dots e \dots c_k}{}_{d_1 \dots d_l} \nonumber \\
&& - \sum^{l}_{j=1} \widehat{R}_{abd_j}{}^{e} A^{c_1 \dots c_k}{}_{d_1 \dots e \dots d_l}. \label{curvature_def}
\end{eqnarray}
The Ricci tensor is introduced by $\widehat{R}_{ij} := \widehat{R}_{kij}{}^k$, and the curvature scalar is $\widehat{R} := g^{ij}\widehat{R}_{ij}$. The signature of the spacetime metric is assumed to be $(+1,-1,-1,-1)$.

In the derivation of the equations of motion we made use of the bitensor formalism; see, e.g., \cite{Synge:1960,DeWitt:Brehme:1960,Poisson:etal:2011} for introductions and references. In particular, the world function is defined as an integral $\sigma(x,y) := \frac{1}{2} \epsilon \left( \int\limits_x^y d\tau \right)^2$ over the geodesic curve connecting the spacetime points $x$ and $y$, where $\epsilon = \pm 1$ for timelike/spacelike curves. Note that our curvature conventions differ from those in \cite{Synge:1960,Poisson:etal:2011}. Indices attached to the world function always denote covariant derivatives, at the given point, i.e.\ $\sigma_y:= \nabla_y \sigma$; hence, we do not make explicit use of the semicolon in the case of the world function. The parallel propagator by $g^y{}_x(x,y)$ allows for the parallel transportation of objects along the unique geodesic that links the points $x$ and $y$. For example, given a vector $V^x$ at $x$, the corresponding vector at $y$ is obtained by means of the parallel transport along the geodesic curve as $V^y = g^y{}_x(x,y)V^x$. For more details see, e.g., section 5 in \cite{Poisson:etal:2011}. A compact summary of useful formulas in the context of the bitensor formalism can also be found in the appendices A and B of \cite{Puetzfeld:Obukhov:2013:1}.

We start by stating, without proof, the following useful rule for a bitensor $B$ with arbitrary indices at different points (here just denoted by dots):
\begin{eqnarray}
\left[B_{\dots} \right]_{;y} = \left[B_{\dots ; y} \right] + \left[B_{\dots ; x} \right]. \label{synges_rule}
\end{eqnarray}
Here a coincidence limit of a bitensor $B_{\dots}(x,y)$ is a tensor 
\begin{eqnarray}
\left[B_{\dots} \right] = \lim\limits_{x\rightarrow y}\,B_{\dots}(x,y),\label{coin}
\end{eqnarray}
determined at $y$. Furthermore, we collect the following useful identities: 
\begin{eqnarray}
&&\sigma_{y_0 y_1 x_0 y_2 x_1} = \sigma_{y_0 y_1 y_2 x_0 x_1} = \sigma_{x_0 x_1 y_0 y_1 y_2 }, \label{rule_1} \\
&&g^{x_1 x_2} \sigma_{x_1} \sigma_{x_2} = 2 \sigma = g^{y_1 y_2} \sigma_{y_1} \sigma_{y_2}, \label{rule_2}\\
&&\left[ \sigma \right]=0, \quad  \left[ \sigma_x \right] = \left[ \sigma_y \right]  = 0, \label{rule_3} \\
&& \left[ \sigma_{x_1 x_2} \right] =  \left[ \sigma_{y_1 y_2} \right] = g_{y_1 y_2}, \label{rule_4}\\ 
&& \left[ \sigma_{x_1 y_2} \right] =  \left[ \sigma_{y_1 x_2} \right] = - g_{y_1 y_2}, \label{rule_5}\\ 
&& \left[ \sigma_{x_1 x_2 x_3} \right] = \left[ \sigma_{x_1 x_2 y_3} \right] = \left[ \sigma_{x_1 y_2 y_3} \right] = \left[ \sigma_{y_1 y_2 y_3} \right] = 0, \nonumber \\ \label{rule_6}\\
&&\left[g^{x_0}{}_{y_1} \right] = \delta^{y_0}{}_{y_1}, \quad \left[g^{x_0}{}_{y_1 ; x_2} \right] = \left[g^{x_0}{}_{y_1 ; y_2} \right] = 0, \label{rule_7} \\
&& \left[g^{x_0}{}_{y_1 ; x_2 x_3} \right] = \frac{1}{2} \widehat{R}{}^{y_0}{}_{y_1 y_2 y_3}. \label{rule_8}
\end{eqnarray}

\section{Covariant expansions}\label{expansion_app}

Here we briefly summarize the covariant expansions of the second derivative of the world function, and the derivative of the parallel propagator:
\begin{eqnarray}
\sigma^{y_0}{}_{x_1} &=& g^{y'}{}_{x_1}\biggl( -\,\delta^{y_0}{}_{y'}\nonumber\\
&& +\,\sum\limits_{k=2}^\infty\,{\frac {1}{k!}}\,\alpha^{y_0}{}_{y'y_2\!\dots \!y_{k+1}}\sigma^{y_2}\cdots\sigma^{y_{k+1}}\biggr)\!,\label{app_expansion_1}\\
\sigma^{y_0}{}_{y_1} &=& \delta^{y_0}{}_{y_1} \nonumber\\
&& -\,\sum\limits_{k=2}^\infty\,{\frac {1}{k!}}\,\beta^{y_0}{}_{y_1y_2\dots y_{k+1}} \sigma^{y_2}\!\cdots\!\sigma^{y_{k+1}}, \label{app_expansion_2} \\
g^{y_0}{}_{x_1 ; x_2} &=& g^{y'}{\!}_{x_1} g^{y''}{\!}_{x_2}\biggl({\frac 12} 
\widehat{R}{}^{y_0}{}_{y'y''y_3}\sigma^{y_3}\nonumber\\ 
&&\!+\!\sum\limits_{k=2}^\infty\,{\frac {1}{k!}}\,\gamma^{y_0}{}_{y'y''y_3\dots y_{k+2}}\sigma^{y_3}\!\cdots\!\sigma^{y_{k+2}}\!\biggr)\!,\label{app_expansion_3} \\
g^{y_0}{}_{x_1 ; y_2} &=& g^{y'}{\!}_{x_1} \biggl({\frac 12} \widehat{R}{}^{y_0}{}_{y'y_2y_3}\sigma^{y_3}\nonumber\\ 
&&\!+\!\sum\limits_{k=2}^\infty\,{\frac {1}{k!}}\,\gamma^{y_0}{}_{y'y_2y_3\dots y_{k+2}}\sigma^{y_3}\!\cdots\!\sigma^{y_{k+2}}\!\biggr).\label{app_expansion_4}\\
G^{Y_0}{}_{X_1 ; x_2} &=& G^{Y'}{\!}_{X_1} g^{y''}{\!}_{x_2} \sum\limits_{k=1}^\infty\,{\frac {1}{k!}}\,\gamma^{Y_0}{}_{Y'y''y_3\dots y_{k+2}}\sigma^{y_3}\!\cdots\!\sigma^{y_{k+2}},\nonumber\\
&& \label{app_expansion_5} \\
G^{Y_0}{}_{X_1 ; y_2} &=& G^{Y'}{\!}_{X_1} \sum\limits_{k=1}^\infty\,{\frac {1}{k!}}\,\gamma^{Y_0}{}_{Y'y_2y_3\dots y_{k+2}}\sigma^{y_3}\!\cdots\!\sigma^{y_{k+2}}.\label{app_expansion_6}
\end{eqnarray}
The coefficients $\alpha, \beta, \gamma$ in these expansions are polynomials constructed from the Riemann curvature tensor and its covariant derivatives. The first coefficients read as follows:
\begin{eqnarray}
\alpha^{y_0}{}_{y_1y_2y_3} &=& - \frac{1}{3} \widehat{R}{}^{y_0}{}_{(y_2y_3)y_1},\label{a1}\\
\beta^{y_0}{}_{y_1y_2y_3} &=& \frac{2}{3}\widehat{R}{}^{y_0}{}_{(y_2y_3)y_1},\label{be1}\\
\alpha^{y_0}{}_{y_1y_2y_3y_4} &=& \frac{1}{2} \widehat{\nabla}_{(y_2}\widehat{R}{}^{y_0}{}_{y_3y_4)y_1},\label{al2}\\
\beta^{y_0}{}_{y_1y_2y_3y_4} &=& - \frac{1}{2} \widehat{\nabla}_{(y_2} \widehat{R}{}^{y_0}{}_{y_3y_4)y_1},\label{be2}\\
\nonumber\\
\gamma^{y_0}{}_{y_1y_2y_3y_4}&=& \frac{1}{3} \widehat{\nabla}_{(y_3} \widehat{R}{}^{y_0}{}_{|y_1|y_4)y_2}.\label{ga}
\end{eqnarray}
We also need the covariant expansion of a usual vector:
\begin{eqnarray} 
A_x = g^{y_0}{}_x\,\sum\limits_{k=0}^\infty\,{\frac {(-1)^k}{k!}} \, A_{y_0;y_1\dots y_k}\,\sigma^{y_1}\cdots\sigma^{y_k}.\label{Ax}
\end{eqnarray}

\section{Explicit form}\label{explicit_app}

Here we make contact with our notation in \cite{Puetzfeld:Obukhov:2013:3} to facilitate a direct comparison to the results there. 

We introduce the auxiliary variables 
\begin{eqnarray}
\Phi^{y_1\dots y_ny_0}{}_{x_0} &:=& \sigma^{y_1} \cdots \sigma^{y_n} g^{y_0}{}_{x_0},\label{Phi}\\
\Psi^{y_1\dots y_ny_0y'}{}_{x_0x'} &:=& \sigma^{y_1} \cdots \sigma^{y_n} g^{y_0}{}_{x_0}g^{y'}{}_{x'}.
\label{Psi}
\end{eqnarray}
\begin{widetext}
Their derivatives 
\begin{eqnarray}
\Psi^{y_1\dots y_ny_0y'}{}_{x_0x';z} &=& \sum^{n}_{a=1}\sigma^{y_1}\cdots\sigma^{y_a}{}_z\cdots\sigma^{y_n}g^{y_0}{}_{x_0}g^{y'}{}_{x'} 
+ \sigma^{y_1} \cdots \sigma^{y_n}\left(g^{y_0}{}_{x_0;z}g^{y'}{}_{x'} + g^{y_0}{}_{x_0}g^{y'}{}_{x';z}\right)\label{dPsi},\\
\Phi^{y_1\dots y_ny_0}{}_{x_0;z} &=& \sum^{n}_{a=1}\sigma^{y_1}\cdots\sigma^{y_a}{}_z\cdots\sigma^{y_n}g^{y_0}{}_{x_0}
+ \sigma^{y_1} \cdots \sigma^{y_n}\,g^{y_0}{}_{x_0;z},\label{dPhi}
\end{eqnarray}
can be straightforwardly evaluated by using the expansion from appendix \ref{expansion_app}. 

In terms of (\ref{Phi}) and (\ref{Psi}) the integrated conservation laws (\ref{cons1f}) and (\ref{cons2f}) take the form 
\begin{eqnarray}
&&{\frac{D}{ds}} \int \Psi^{y_1\dots y_ny_0y'}{}_{x_0x'}\widetilde{\Delta}^{x_0 x' x_2} d\Sigma_{x_2} = 
\int \Psi^{y_1\dots y_ny_0y'}{}_{x_0x'}\left[ - U_{x''''}{}^{x'}{}_{x''}{}^{x_0}{}_{x'''}\widetilde{\Delta}^{x'' x''' x''''} 
+ \widetilde{\Sigma}^{x' x_0} - \widetilde{t}^{x' x_0}\right] w^{x_2} d\Sigma_{x_2}\nonumber\\  
&& +\,\int\Psi^{y_1\dots y_ny_0y'}{}_{x_0x';x''}\widetilde{\Delta}^{x_0 x' x''}w^{x_2} d\Sigma_{x_2} 
+ \int v^{y_{n+1}}\Psi^{y_1\dots y_ny_0y'}{}_{x_0x';y_{n+1}}\widetilde{\Delta}^{x_0 x' x_2}d\Sigma_{x_2}, \label{int_eom_1}
\end{eqnarray}
\begin{eqnarray}
&&{\frac{D}{ds}} \int\Phi^{y_1\dots y_ny_0}{}_{x_0} \widetilde{\Sigma}^{x_0 x_2} d\Sigma_{x_2} = \int \Phi^{y_1\dots y_ny_0}{}_{x_0} \left(-V_{x''}{}^{x_0}{}_{x'} \widetilde{\Sigma}^{x' x''} - R^{x_0}{}_{x''' x' x''} \widetilde{\Delta}^{x' x'' x'''}
- \frac{1}{2} Q^{x_0}{}_{x'' x'} \widetilde{t}^{x' x''} \right.\nonumber\\
&&\left. -\,A^{x_0} \widetilde{L}_{\rm mat}\right) w^{x_2} d\Sigma_{x_2} 
+ \int\Phi^{y_1\dots y_ny_0}{}_{x_0;x'}\widetilde{\Sigma}^{x_0 x'}w^{x_2} d\Sigma_{x_2} 
+ \int v^{y_{n+1}}\Phi^{y_1\dots y_ny_0}{}_{x_0;y_{n+1}} \widetilde{\Sigma}^{x_0 x_2} d\Sigma_{x_2}.\label{int_eom_2}
\end{eqnarray}
This form allows for a direct comparison to (29) and (30) in \cite{Puetzfeld:Obukhov:2013:3}. Explicitly, in terms of (\ref{Phi}) and (\ref{Psi}) the integrated moments from (\ref{jMAG})--(\ref{mMAG}) are given by
\begin{eqnarray}
p^{y_1\dots y_n y_0} &:=& (-1)^n\int\limits_{\Sigma(\tau)}\Phi^{y_1\dots y_n y_0}{}_{x_0} \widetilde{\Sigma}^{x_0 x_1}d\Sigma_{x_1},\\
k^{y_2\dots y_{n+1} y_0 y_1} &:=& (-1)^n\int\limits_{\Sigma(\tau)}\Psi^{{y_2}\dots{y_{n+1} y_0 y_1}}{}_{x_0 x_1}\widetilde{\Sigma}^{x_0 x_1}w^{x_2}d\Sigma_{x_2},\\
h^{y_2\dots y_{n+1}y_0 y_1} &:=& (-1)^n\int\limits_{\Sigma(\tau)}\Psi^{y_2 \dots y_{n+1}y_0 y_1}{}_{x_0 x_1 }\widetilde{\Delta}^{x_0 x_1 x_2}d\Sigma_{x_2},\label{Smom}\\
q^{y_3\dots y_{n+2}y_0 y_1 y_2} &:=& (-1)^n\int\limits_{\Sigma(\tau)}\Psi^{y_3 \dots y_{n+2} y_0 y_1}{}_{x_0 x_1} g^{y_2}{}_{x_2} \widetilde{\Delta}^{x_0 x_1 x_2 }w^{x_3}d\Sigma_{x_3},\\
\mu^{y_2\dots y_{n+1} y_0 y_1} &:=& (-1)^n\int\limits_{\Sigma(\tau)}\Psi^{y_2 \dots y_{n+1} y_0 y_1}{}_{x_0 x_1}\widetilde{t}^{x_0 x_1}w^{x_2}d\Sigma_{x_2},\\
\xi^{y_1\dots y_{n}} &:=& (-1)^n\int\limits_{\Sigma(\tau)}\sigma^{y_1}\cdots\sigma^{y_{n}}L_{\rm mat}w^{x_2}d\Sigma_{x_2}.
\end{eqnarray}
\end{widetext}

\bibliographystyle{unsrtnat}
\bibliography{eommag}

\begin{thebibliography}{36}
\providecommand{\natexlab}[1]{#1}
\providecommand{\url}[1]{\texttt{#1}}
\expandafter\ifx\csname urlstyle\endcsname\relax
  \providecommand{\doi}[1]{doi: #1}\else
  \providecommand{\doi}{doi: \begingroup \urlstyle{rm}\Url}\fi

\bibitem[{Hehl} et~al.(1995){Hehl}, {McCrea}, {Mielke}, and
  {Ne'eman}]{Hehl:1995}
F.~W. {Hehl}, J.~D. {McCrea}, E.~W. {Mielke}, and Y.~{Ne'eman}.
\newblock {Metric-affine gauge theory of gravity: Field equations, Noether
  identities, world spinors, and breaking of dilation invariance}.
\newblock \emph{Phys. Rep.}, 258:\penalty0 1, 1995.

\bibitem[{Blagojevi\'c}(2002)]{Blagojevic:2002}
M.~{Blagojevi\'c}.
\newblock \emph{{Gravitation and Gauge Symmetries}}.
\newblock IOP Publishing, London, 2002.

\bibitem[{Blagojevi\'c} and {Hehl}(2013)]{Hehl:2013}
M.~{Blagojevi\'c} and F.~W. {Hehl}.
\newblock \emph{{Gauge Theories of Gravitation. A Reader with Commentaries}}.
\newblock Imperial College Press, London, 2013.

\bibitem[Mathisson(1937)]{Mathisson:1937}
M.~Mathisson.
\newblock {Neue Mechanik materieller Systeme}.
\newblock \emph{Acta Phys. Pol.}, 6:\penalty0 163, 1937.

\bibitem[Papapetrou(1951)]{Papapetrou:1951:3}
A.~Papapetrou.
\newblock {Spinning test-particles in General Relativity. I}.
\newblock \emph{Proc. Roy. Soc. Lond. A}, 209:\penalty0 248, 1951.

\bibitem[{Dixon}(1964)]{Dixon:1964}
W.~G. {Dixon}.
\newblock {A covariant multipole formalism for extended test bodies in General
  Relativity}.
\newblock \emph{Nuovo Cimento}, 34:\penalty0 317, 1964.

\bibitem[{Dixon}(1974)]{Dixon:1974}
W.~G. {Dixon}.
\newblock {Dynamics of extended bodies in General Relativity. III. Equations of
  motion}.
\newblock \emph{Phil. Trans. R. Soc. Lond. A}, 277:\penalty0 59, 1974.

\bibitem[{Dixon}(1979)]{Dixon:1979}
W.~G. {Dixon}.
\newblock {Extended bodies in General Relativity: Their description and
  motion}.
\newblock \emph{Proc. Int. School of Phys. Enrico Fermi LXVII, Ed. J. Ehlers,
  North Holland, Amsterdam}, page 156, 1979.

\bibitem[{Dixon}(2008)]{Dixon:2008}
W.~G. {Dixon}.
\newblock {Mathisson's new mechanics: Its aims and realisation}.
\newblock \emph{Acta Phys. Pol. B Proc. Suppl.}, 1:\penalty0 27, 2008.

\bibitem[{Synge}(1960)]{Synge:1960}
J.~L. {Synge}.
\newblock \emph{{Relativity: The general theory}}.
\newblock North-Holland, Amsterdam, 1960.

\bibitem[{DeWitt} and {Brehme}(1960)]{DeWitt:Brehme:1960}
B.~S. {DeWitt} and R.~W. {Brehme}.
\newblock {Radiation damping in a gravitational field}.
\newblock \emph{Ann. Phys (N.Y.)}, 9:\penalty0 220, 1960.

\bibitem[{Bertolami} et~al.(2007){Bertolami}, {B\"ohmer}, {Harko}, and
  {Lobo}]{Bertolami:etal:2007}
O.~{Bertolami}, C.~G. {B\"ohmer}, T.~{Harko}, and F.~S.~N. {Lobo}.
\newblock {Extra force in $f(R)$ modified theories of gravity}.
\newblock \emph{Phys. Rev. D.}, 75:\penalty0 104016, 2007.

\bibitem[{Nojiri} and {Odintsov}(2011)]{Nojiri:2011}
S.~{Nojiri} and S.~D. {Odintsov}.
\newblock {Unified cosmic history in modified gravity: from $F(R)$ theory to
  Lorentz non-invariant models}.
\newblock \emph{Phys. Rep.}, 505:\penalty0 59, 2011.

\bibitem[{Straumann}(2008)]{Straumann:2008}
N.~{Straumann}.
\newblock {Problems with Modified Theories of Gravity, as Alternatives to Dark
  Energy}.
\newblock 2008.
\newblock URL \url{arXiv:0809.5148v1 [gr-qc]}.

\bibitem[{Harko}(2014)]{Harko:2014:1}
T.~{Harko}.
\newblock {Thermodynamic interpretation of the generalized gravity models with
  geometry - matter coupling}.
\newblock \emph{Phys. Rev. D}, 90:\penalty0 044067, 2014.

\bibitem[{Stoeger} and {Yasskin}(1979)]{Stoeger:Yasskin:1979}
W.~R. {Stoeger} and P.~B. {Yasskin}.
\newblock {Can a macroscopic gyroscope feel torsion?}
\newblock \emph{Gen. Rel. Grav.}, 11:\penalty0 427, 1979.

\bibitem[{Yasskin} and {Stoeger}(1980)]{Stoeger:Yasskin:1980}
P.~B. {Yasskin} and W.~R. {Stoeger}.
\newblock {Propagation equations for test bodies with spin and rotation in
  theories of gravity with torsion}.
\newblock \emph{Phys. Rev. D}, 21:\penalty0 2081, 1980.

\bibitem[{Puetzfeld} and {Obukhov}(2007)]{Puetzfeld:2007}
D.~{Puetzfeld} and Yu.N. {Obukhov}.
\newblock {Propagation equations for deformable test bodies with microstructure
  in extended theories of gravity}.
\newblock \emph{Phys. Rev. D}, 76:\penalty0 084025, 2007.

\bibitem[{Puetzfeld} and
  {Obukhov}(2008{\natexlab{a}})]{Puetzfeld:Obukhov:2008:1}
D.~{Puetzfeld} and Yu.~N. {Obukhov}.
\newblock {Motion of test bodies in theories with nonminimal coupling}.
\newblock \emph{Phys. Rev. D}, 78:\penalty0 121501, 2008{\natexlab{a}}.

\bibitem[{Puetzfeld} and
  {Obukhov}(2008{\natexlab{b}})]{Puetzfeld:Obukhov:2008:2}
D.~{Puetzfeld} and Yu.~N. {Obukhov}.
\newblock {Probing non-Riemannian spacetime geometry}.
\newblock \emph{Phys. Lett. A}, 372:\penalty0 6711, 2008{\natexlab{b}}.

\bibitem[{Puetzfeld} and
  {Obukhov}(2013{\natexlab{a}})]{Puetzfeld:Obukhov:2013:1}
D.~{Puetzfeld} and Yu.~N. {Obukhov}.
\newblock {Covariant equations of motion for test bodies in gravitational
  theories with general nonminimal coupling}.
\newblock \emph{Phys. Rev. D}, 87:\penalty0 044045, 2013{\natexlab{a}}.

\bibitem[{Hehl} et~al.(2013){Hehl}, {Obukhov}, and
  {Puetzfeld}]{Hehl:Obukhov:Puetzfeld:2013}
F.~W. {Hehl}, Yu.~N. {Obukhov}, and D.~{Puetzfeld}.
\newblock {On Poincar\'e gauge theory of gravity, its equations of motion, and
  Gravity Probe B}.
\newblock \emph{Phys. Lett. A}, 377:\penalty0 1775, 2013.

\bibitem[{Puetzfeld} and
  {Obukhov}(2013{\natexlab{b}})]{Puetzfeld:Obukhov:2013:3}
D.~{Puetzfeld} and Yu.~N. {Obukhov}.
\newblock {Equations of motion in gravity theories with nonminimal coupling: A
  loophole to detect torsion macroscopically?}
\newblock \emph{Phys. Rev. D}, 88:\penalty0 064025, 2013{\natexlab{b}}.

\bibitem[{Puetzfeld} and
  {Obukhov}(2013{\natexlab{c}})]{Puetzfeld:Obukhov:2013:4}
D.~{Puetzfeld} and Yu.~N. {Obukhov}.
\newblock {Unraveling gravity beyond Einstein with extended test bodies}.
\newblock \emph{Phys. Lett. A}, 377:\penalty0 2447, 2013{\natexlab{c}}.

\bibitem[{Roshan}(2013)]{Roshan:2013}
M.~{Roshan}.
\newblock {Test particle motion in modified gravity theories}.
\newblock \emph{Phys. Rev. D}, 87:\penalty0 044005, 2013.

\bibitem[{Puetzfeld} and {Obukhov}(2014)]{Puetzfeld:Obukhov:2014:1}
D.~{Puetzfeld} and Yu.~N. {Obukhov}.
\newblock {Prospects of detecting spacetime torsion}.
\newblock \emph{Int. J. Mod. Phys. D}, 23:\penalty0 1442004, 2014.

\bibitem[{Obukhov} and {Tresguerres}(1993)]{Obukhov:1993}
Yu.~N. {Obukhov} and R.~{Tresguerres}.
\newblock {Hyperfluid - a model of classical matter with hypermomentum}.
\newblock \emph{Phys. Lett. A}, 184:\penalty0 17, 1993.

\bibitem[{Taub}(1954)]{Taub:1954}
A.~H. {Taub}.
\newblock {General relativistic variational principle for perfect fluids}.
\newblock \emph{Phys. Rev.}, 94:\penalty0 1468, 1954.

\bibitem[{Schutz}(1970)]{Schutz:1970}
B.~F. {Schutz}.
\newblock {Perfect fluids in general relativity: Velocity potentials and
  variational principles}.
\newblock \emph{Phys. Rev. D}, 2:\penalty0 2762, 1970.

\bibitem[{Weyssenhoff} and {Raabe}(1947)]{Weyssenhoff:1947}
J.~{Weyssenhoff} and A.~{Raabe}.
\newblock {Relativistic dynamics of spin-fluids and spin-particles}.
\newblock \emph{Acta Phys. Pol.}, 9:\penalty0 7, 1947.

\bibitem[{Obukhov} and {Korotky}(1987)]{Obukhov:1987}
Yu.~N. {Obukhov} and V.~A. {Korotky}.
\newblock {The Weyssenhoff fluid in Einstein-Cartan theory}.
\newblock \emph{Class. Quantum Grav.}, 4:\penalty0 1633, 1987.

\bibitem[{Obukhov} and {Puetzfeld}(2014)]{Obukhov:Puetzfeld:2014}
Yu.~N. {Obukhov} and D.~{Puetzfeld}.
\newblock {Conservation laws in gravity: A unified framework}.
\newblock \emph{Phys. Rev. D}, 90:\penalty0 024004, 2014.

\bibitem[{Dixon}(1967)]{Dixon:1967}
W.~G. {Dixon}.
\newblock {Description of extended bodies by multipole moments in special
  relativity}.
\newblock \emph{J. Math. Phys.}, 8:\penalty0 1591, 1967.

\bibitem[{Weyl}(1923)]{Weyl:1923}
H.~{Weyl}.
\newblock \emph{{Raum-Zeit-Materie}}.
\newblock Springer, Berlin, 1923.

\bibitem[{Bailey} and {Israel}(1975)]{Bailey:Israel:1975}
I.~{Bailey} and W.~{Israel}.
\newblock {Lagrangian dynamics of spinning particles and polarized media in
  General Relativity}.
\newblock \emph{Comm. Math. Phys.}, 42:\penalty0 65, 1975.

\bibitem[{Poisson} et~al.(2011){Poisson}, {Pound}, and
  {Vega}]{Poisson:etal:2011}
E.~{Poisson}, A.~{Pound}, and I.~{Vega}.
\newblock {The motion of point particles in curved spacetime}.
\newblock \emph{Living Reviews in Relativity}, 14\penalty0 (7), 2011.

\end{thebibliography}

\end{document}